\documentclass[aps,prd,twocolumn,groupedaddress,showpacs]{revtex4}

\usepackage{graphicx}
\usepackage{dcolumn}
\usepackage{bm}
\usepackage{amsmath}
\usepackage{epstopdf}
\usepackage{amsfonts}
\usepackage{amssymb}
\usepackage{hyperref}
\usepackage{xcolor}

\usepackage{natbib}

\providecommand{\U}[1]{\protect\rule{.1in}{.1in}}

\newcommand{\baa}{\begin{align}}
\newcommand{\eaa}{\end{align}}

\newcommand{\be}{\begin{equation}}
\newcommand{\ee}{\end{equation}}
\newcommand{\bea}{\begin{eqnarray}}
\newcommand{\eea}{\end{eqnarray}}

\begin{document}



\title{Quasinormal modes of scale dependent black holes \\
in (1+2)-dimensional Einstein-power-Maxwell theory}


\author{\'Angel Rinc\'on}
\affiliation{Instituto de F\'{i}sica, Pontificia Universidad Cat\'{o}lica de Chile, \mbox{Avenida Vicu\~na Mackenna 4860, Santiago, Chile.}}
\email{arrincon@uc.cl}

\author{Grigoris Panotopoulos}
\affiliation{Centro Multidisciplinar de Astrof\'{\i}sica, Instituto Superior T\'ecnico,
Universidade de Lisboa, Av. Rovisco Pais, 1049-001 Lisboa, Portugal}
\email{grigorios.panotopoulos@tecnico.ulisboa.pt}

%

\date{\today}

\begin{abstract}
We study for the first time the stability against scalar perturbations, and we compute the spectrum of quasinormal modes of three-dimensional charged black holes in Einstein-power-Maxwell non-linear electrodynamics assuming running couplings. Adopting the 6th order WKB approximation we investigate how the running of the couplings change the spectrum of the classical theory. Our results show that all modes corresponding to non-vanishing angular momentum are unstable both in the classical theory and with the running of the couplings, while the fundamental mode can be stable or unstable depending on the running parameter and the electric charge.
\end{abstract}

\maketitle


\section{Introduction}

A consistent formulation of quantum gravity is still an open task. Although so far there are several approaches to quantum gravity
in the literature (see e.g. \cite{QG1,QG2,QG3,QG4,QG5,QG6,QG7,QG8,QG9} and references therein), most of them share a common property, namely
that the basic parameters that enter into the action, such as Newton's constant, the electromagnetic coupling or the cosmological constant, become scale dependent quantities. This does not come as a surprise, since scale dependence at the level of the effective action is a generic result of quantum field theory. Scale dependent couplings are expected to modify, and indeed they do, the properties of classical black hole backgrounds (see section 2 below).

Black holes (BHs), a generic prediction of Einstein's General Relativity (GR), are way more than just mathematical objects. After Hawking's seminal work \cite{hawking1,hawking2} in which it was shown that BHs emit radiation from their horizon, these objects have become of paramount importance in theories of gravity, and an excellent laboratory to understand quantum gravity. Greybody factors and quasinormal modes are of special interest that have attracted a lot of attention over the last years. First, greybody factors are frequency dependent quantities that measure the modification of the original black body radiation, since the emitted particles feel an effective potential barrier that backscatters a part of the outcoming radiation back into the black hole. On the other hand, (in)stability of a system and how it responds to a small external perturbation have been always important issues in physics. In particular, stability of BHs against small perturbations is an old subject started with the works of \cite{wheeler,zerilli1,zerilli2,zerilli3,moncrief,teukolsky} (see also Chandrasekhar's monograph \cite{monograph}). Quasinormal modes (QNM) with a non-vanishing imaginary part, depend entirely on the few BH parameters, and thus they contain unique information about the mass, electric charge and angular momentum of black holes. After the LIGO direct detections of gravitational waves \cite{ligo1,ligo2,ligo3}, that offer us the strongest evidence so far that BH exist and merge, QNM of black holes are more relevant than ever. By observing the quasinormal spectrum, that is frequencies and damping rates, we can determine the black hole parameters. Although greybody factors and quasinormal modes at first sight may seem completely unrelated, they are in fact closely related and differ only by the boundary conditions of the same mathematical problem, see section 3. For a review on BH QNM see \cite{review1}, and for a more recent one \cite{review2}.

Gravity in (1+2) dimensions, mainly due to the absence of propagating degrees of freedom as well as its deep connection to a Yang-Mills theory with only the Chern-Simons term \cite{CS,witten1,witten2}, is definitely special and allows us to study BH that share properties of their four-dimensional counterparts, such as Hawking radiation and thermodynamical properties, in a simpler mathematical framework. In the original BTZ black hole \cite{btz1,btz2} the presence of a cosmological constant was crucial for the existence of the BH horizon. If, however, the black hole is electrically charged there is a horizon even without the cosmological constant. Standard Maxwell's electrodynamics in four dimensions is both linear and characterized by a traceless energy momentum tensor. In (1+2)-dimensional spacetimes, however, in the linear theory the trace of the stress energy tensor does not vanish any more. It is straightforward to generalize the theory by assuming a non-linear electrodynamics described by the Lagrangian density $\mathcal{L} = F^k$, where
$F$ is Maxwell's invariant and $k$ is an arbitrary rational number. In this class of theories, called Einstein-power-Maxwell theories (EpM), the stress energy tensor is traceless for $k=D/4$, where $D$ is the dimensionality of spacetime. Therefore, if $k=3/4$ we have a non-linear theory in 3 dimensions with vanishing trace of the energy momentum tensor. Black hole solutions in 3 and higher dimensions have been obtained in \cite{BH1,BH2}, while the running of couplings either in BTZ or in EpM theory in (1+2) gravity has been investigated in \cite{Rincon:2017ypd,Rincon:2017goj}.

It is the aim of the present work to study for the first time the stability and compute the quasinormal spectrum of charged black holes in three-dimensional EpM non-linear electrodynamics.
Our work is organized as follows: After this introduction, we present the model and the BH solution in section 2, while in the third section the effective potential for scalar perturbations is presented. In section 4 we obtain the quasinormal modes, and finally we conclude our work in the last section. We use natural units where $c=1=\hbar$
and metric signature $(-,+,+)$.

\section{Scale dependent black hole solution in EpM theory}
\noindent
In this section we briefly summarize the model, the equations of motion and the solutions both for the classical and
the scale dependent EpM theory. The notation follows closely that of previous works on the subject
\cite{Rincon:2017ypd,Rincon:2017goj,Koch:2014joa,angel,Rincon:2017ayr}. 

\subsection{Classical EpM}

First we consider the classical action
\begin{align}\label{Classical_Action}
S_0 & =  \int {\mathrm {d}}^{3}x {\sqrt {-g}}\,
\bigg[\frac{1}{2 \kappa_0} R - \frac{1}{e_{0}^{2k}}\mathcal{L}(F) \bigg],
\end{align}
where $R$ is the Ricci scalar, $G_0=\kappa_0/8\pi$ and $e_0$ are the gravitational and electromagnetic couplings respectively, $F=(1/4)F_{\mu \nu} F^{\mu \nu}$ is the Maxwell invariant, $F_{\mu \nu}=\partial_\mu A_\nu-\partial_\nu A_\mu$ is the electromagnetic field strength tensor, and $k$ is the arbitrary rational number that defines the theory. Varying with respect to the Maxwell field $A_\mu$ we obtain its equation of motion
\begin{equation}
D_\mu \left ( \frac{\mathcal{L}_F F^{\mu \nu}}{e_0^{2k}} \right ) = 0 
\end{equation}
while varying with respect to the metric $g_{\mu \nu}$  we obtain Einstein's field equations
\begin{equation}
G_{\mu \nu} = \frac{\kappa_0}{e_0^{2k}} T_{\mu \nu}^{\text{EM}}
\end{equation}
where $\mathcal{L}_F = \mathrm{d} \mathcal{L}/\mathrm{d} F$ and $T_{\mu \nu}^{\text{EM}}=\mathcal{L}(F) g_{\mu \nu}-\mathcal{L}_F F_{\mu \gamma} F_\nu^\gamma$ is the electromagnetic energy momentum tensor. For circularly symmetric solutions we make the ansatz
\begin{eqnarray}
ds^2 & = & -f_0(r) dt^2 + f_0(r)^{-1} dr^2 + r^2 d \phi^2  \\
F_{\mu \nu} & = & (\delta_\mu^r \delta_\nu^ t- \delta_\nu^r \delta_\mu^t) E_0(r)
\end{eqnarray}
and therefore one has to determine a set of two functions ${f_0(r), E_0(r)}$, that correspond to the metric function and the electric field respectively. Assuming $k=3/4$ the solution is found to be
\begin{eqnarray}
f_0(r) & = & -G_0 M_0 + \frac{4 G_0 Q_0^2}{3r}  \\
E_0(r) & = & \frac{Q_0}{r^2}
\end{eqnarray}
where $M_0$ and $Q_0$ are the classical values of the mass and the electric charge of the BH respectively.
Given the solution it is straightforward to compute some properties of the BH, such as event horizon $r_0$, Hawking temperature
$T_0$ and Bekenstein-Hawking entropy $S_0$, which are found to be
\begin{eqnarray}
r_0 & = & \frac{4 Q_0^2}{3 M_0}  \\
T_0 & = & \frac{M_0 G_0}{4 \pi r_0} \\
S_0 & = & \frac{\pi r_0}{2 G_0}
\end{eqnarray}

\subsection{Scale dependent EpM}

Now we move on to the scale dependent EpM theory, which is described by the action
\begin{align}\label{Effective_Actiom}
\Gamma[g_{\mu \nu}, A_{\mu}, k] &=  \int {\mathrm {d}}^{3}x {\sqrt {-g}}\,
\bigg[\frac{1}{2\kappa_k} R - \frac{1}{e_{k}^{2\beta}}\mathcal{L}(F) \bigg],
\end{align}
Note that now we have the same couplings as before, but they are scale dependent,  $\kappa_k=8 \pi G_k$ and $e_k$. In addition, in this case there three independent fields, namely the metric $g_{\mu \nu}(x)$, the electromagnetic four-potential $A_{\mu}(x)$, and the scale field $k(x)$.

Einstein's field equations as well as the equation of motion for the Maxwell potential maintain their form
\begin{align} \label{decovcoupling}
D_{\mu}\left(\frac{\mathcal{L}_{F}F^{\mu\nu}}{e_k^{2\beta}}\right)& = 0.
\end{align}
and
\begin{align}
G_{\mu\nu} &= \frac{\kappa_k}{e^{2\beta}_k}T^{\text{effec}}_{\mu\nu},
\end{align}
where the couplings are the scale dependent ones, and also the matter energy momentum tensor $T^{\text{effec}}_{\mu\nu}$ is given by
\begin{align}
T^{\text{effec}}_{\mu\nu} &= T^{\text{EM}}_{\mu\nu} - \frac{e^{2\beta}_k}{\kappa_k} \Delta t_{\mu \nu}.
\end{align}
with the additional object $\Delta t_{\mu \nu}$ defined as follows
\begin{align}
\Delta t_{\mu\nu} &= G_k \Bigl(g_{\mu \nu} \square - \nabla_{\mu} \nabla_{\nu}
\Bigl)G_k^{-1}.
\end{align}
At this point a couple of remarks are in order. The renormalization scale $k$ is not a constant, and since there is one consistency equation missing the corresponding system of equations of motion is not a closed one. This implies that the energy momentum tensor is not conserved. Despite that, this kind of problem has been studied, at least at the level of renormalization group improvement of black holes in asymptotic safety scenarios \cite{Bonanno:2000ep,Bonanno:2006eu,Reuter:2006rg,Reuter:2010xb,Falls:2012nd,Cai:2010zh,Becker:2012js,Becker:2012jx,
Koch:2013owa,Koch:2013rwa,Ward:2006vw,Burschil:2009va,Falls:2010he,Koch:2014cqa,Bonanno:2016dyv}.
In order to fix the aforementioned problem, we can use the so--called principle of minimal sensitivity \cite{Reuter:2003ca,Koch:2010nn,Domazet:2012tw,Koch:2014joa,Contreras:2016mdt}. This allows us to obtain an additional equation using the effective action (\ref{Effective_Actiom}) and taking the derivative of it respect to the renormalization scale $k$, i.e.
\begin{equation}\label{vary}
\frac{d}{d k} \Gamma[g_{\mu \nu}, A_{\mu}, k]=0,
\end{equation}
Using Eq. \eqref{vary} and the corresponding equation of motion, we are able to recover the usual energy momentum tensor conservation (for additional details check \cite{Percacci:2016arh} and references therein). A problem is still present: we need to know the explicit form of the the beta function and, in many cases, the precise expression of the beta functions is unknown (or at least unsure). To avoid this problem, we can use additional constraints as was previously reported in Ref. \cite{Rincon:2017goj}, for instance the null energy condition
\begin{equation}
\Delta t_{\mu \nu} l^\mu l^\nu = 0
\end{equation}
with $l^\mu$ being a null vector. With this, we can solve the problem
for the couplings $G(r), e(r)$ etc directly \cite{Contreras:2013hua,Koch:2015nva,angel,Rincon:2017ypd}.
This philosophy of assuring the consistency of the equations by imposing a null energy condition will also be used in the present
work.
Finally, assuming the same ansatz as before for circularly symmetric solutions, the lapse metric function is computed to be
\begin{equation}
f(r) = \frac{4 G_0 Q_0^2}{3 r (1+r \epsilon)^3}-\frac{G_0 M_0 (r^3 \epsilon^2+3 r^2 \epsilon + 3 r)}{3 r (1+r \epsilon)^3}
\end{equation}
where $\epsilon$ is the running parameter which let us move from the classical solution $(\epsilon =0)$ to the scale dependent one $(\epsilon \neq 0)$. With this new metric function the horizon $r_H$, the Hawking temperature $T_H$ as well as the Bekenstein-Hawking entropy are computed to be (at leading order in $\epsilon$)
\begin{eqnarray}
r_H & \simeq & r_0 (1-\epsilon r_0)  \\
T_H & \simeq & T_0 \left (1 + \frac{1}{3}(\epsilon r_0)^2 \right ) \\
S   & \simeq & S_0 \left (1 - \frac{1}{3}(\epsilon r_0)^2 \right )
\end{eqnarray}
which confirms our initial statement that the running of the couplings modify the properties of the classical BH backgrounds.

\section{Scalar perturbations}

\begin{figure}[ht!]
\centering
\includegraphics[width=\linewidth]{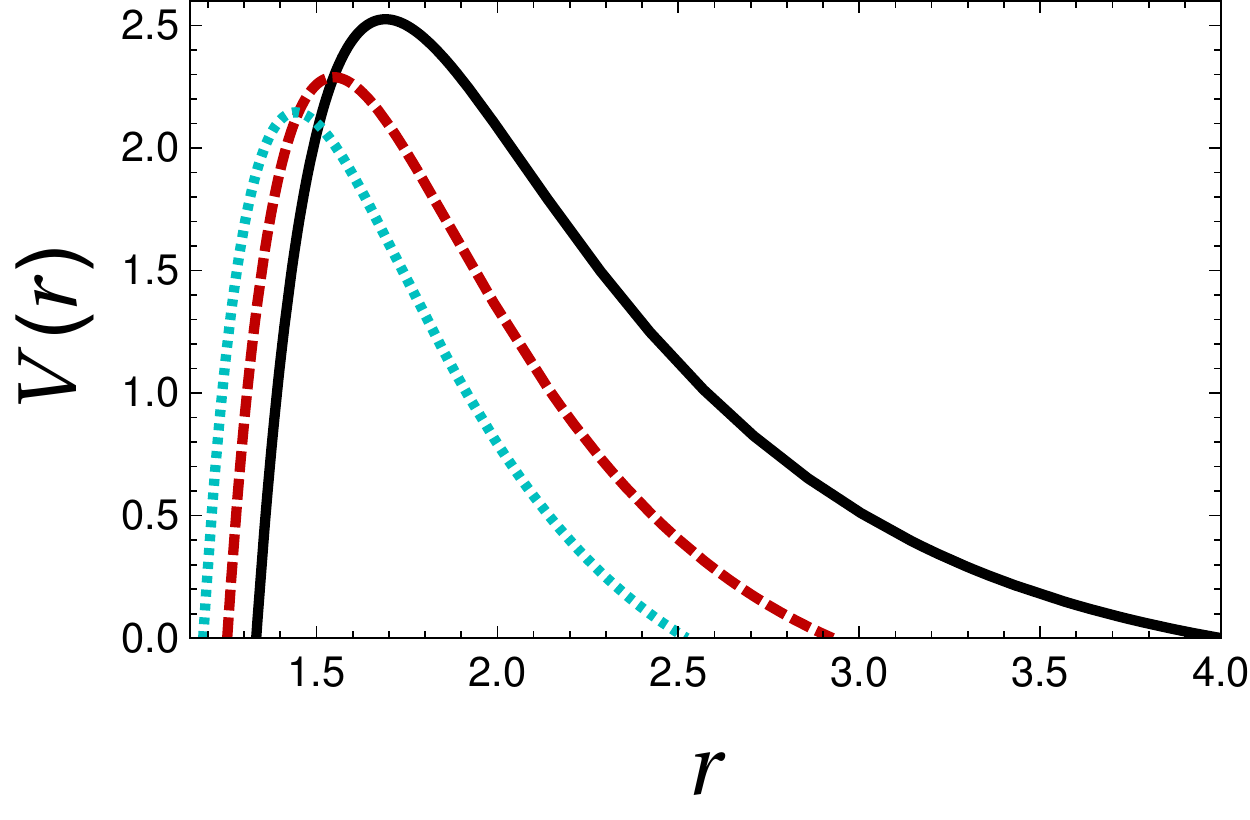}
\caption{\label{fig:1}
Effective potential $V(r)$ taking $G=M=Q=1$ and $m=0$ for $\epsilon=0$, (solid black line), $\epsilon=0.05$ (dashed red line) and $\epsilon=0.1$ (dotted blue line). Note that the vertical axes is scaled to $1:10^2$.
}
\end{figure}

\begin{figure}[ht!]
\centering
\includegraphics[width=\linewidth]{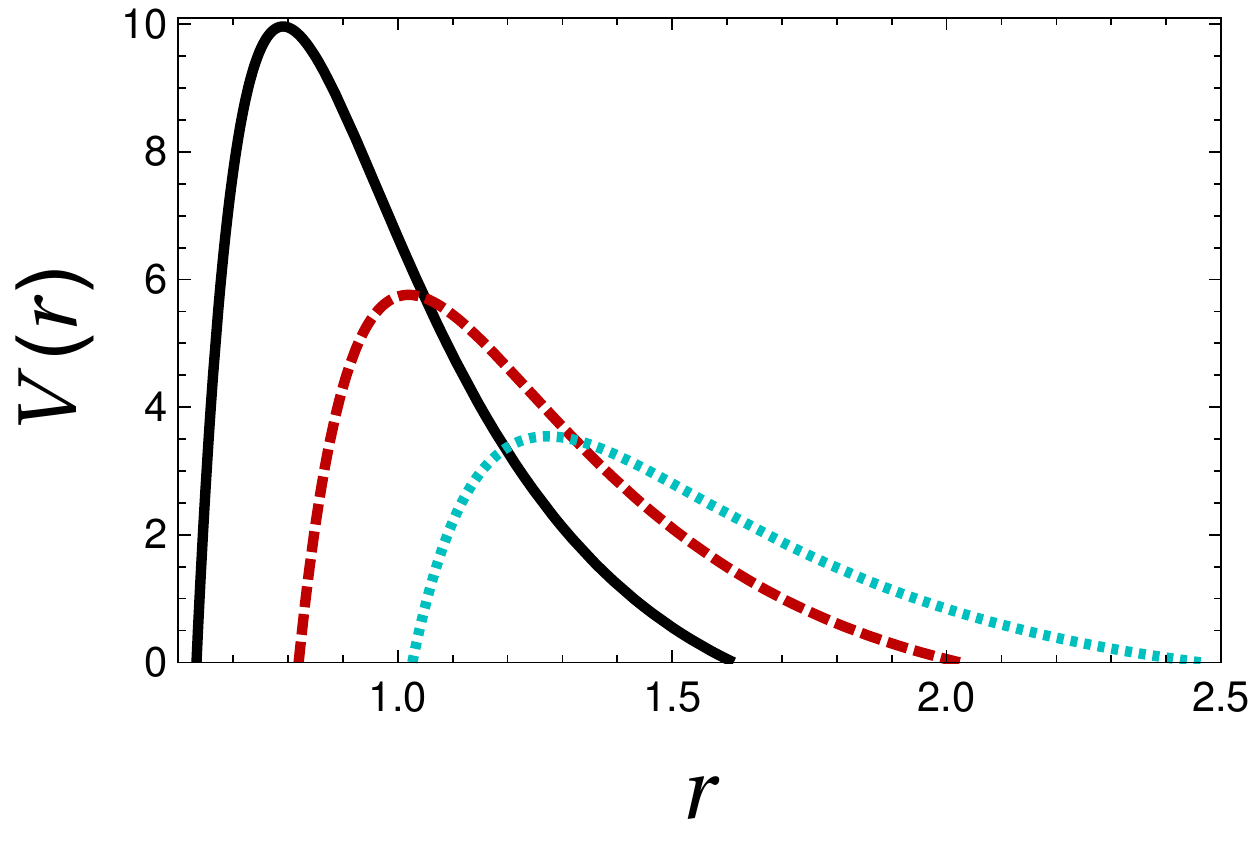}
\caption{\label{fig:2}
Effective potential $V(r)$ taking $G=M=1, \epsilon=0.05$ and $m=0$ for $Q=0.7$, (solid black line), $Q=0.8$ (dashed red line) and $Q=0.9$ (dotted blue line). Note that the vertical axes is scaled to $1:10^2$.
}
\end{figure}

In this section we study the propagation of a probe minimally coupled massless scalar field $\Phi(t,r,\phi)$
in a given gravitational background of the form
\begin{equation}
ds^2 = -f(r) dt^2 + f(r)^{-1} dr^2 + r^2 d \phi^2
\end{equation}
with a known lapse metric function $f(r)$. The starting point is the well-known wave equation
\begin{equation}
\frac{1}{\sqrt{-g}} \partial_\mu (\sqrt{-g} g^{\mu \nu} \partial_\nu) \Phi = 0
\end{equation}
which is a partial differential equation for the scalar field. Next we seek solutions where the time and angular dependence are known as follows
\begin{equation}\label{separable}
\Phi(t,r,\phi) = e^{-i \omega t} R(r) e^{i m \phi}
\end{equation}
with $m$ being is the quantum number of angular momentum. Using the above ansatz it is straightforward to obtain the radial equation, which is an ordinary differential equation
\begin{equation}
R'' + \left( \frac{1}{r} + \frac{f'}{f} \right) R' + \left( \frac{\omega^2}{f^2} - \frac{m^2}{r^2 f} \right) R = 0
\end{equation}
where the prime denotes differentiation with respect to radial distance $r$. Next we recast the equation for the radial part into a Schr{\"o}dinger-like equation of the form
\begin{equation}
\frac{d^2 \psi}{dx^2} + (\omega^2 - V(x)) \psi = 0
\end{equation}
by defining new variables, a dependent $R \rightarrow \psi$ as well as an independent one $r \rightarrow x$ as follows
\begin{eqnarray}
R & = & \frac{\psi}{\sqrt{r}} \\
x & = & \int \frac{dr}{f(r)}
\end{eqnarray}
with $x$ being the so-called tortoise coordinate. Therefore we obtain for the effective potential the expression
\begin{equation}
V(r) = f(r) \: \left( \frac{m^2}{r^2}+\frac{f'(r)}{2 r}-\frac{f(r)}{4 r^2} \right)
\end{equation}
which as a function of the radial coordinate can be seen in Figures \ref{fig:1} (for 3 different values of the running parameter) and  \ref{fig:2} (for 3 different values of the electric charge).

Finally, the Schr{\"o}dinger-like equation must be supplemented by appropriate boundary conditions, which for asymptotically flat spacetimes are the following
\begin{equation}
\psi(x) \rightarrow
\left\{
\begin{array}{lcl}
A e^{i \omega x} & \mbox{ if } & x \rightarrow - \infty \\
&
&
\\
 C_+ e^{i \omega x} + C_- e^{-i \omega x} & \mbox{ if } & x \rightarrow + \infty
\end{array}
\right.
\end{equation}
where $A, C_+, C_-$ are arbitrary constants. Up to now, following the procedure just described one can compute the so-called greybody factors (GBF), which as already mentioned in the introduction show the modification of the spectrum of Hawking radiation due to the effective potential barrier, and where the frequency is real and takes continuous values. For an incomplete list see e.g. \cite{col1,col2,col3,col4,col5,col6,col7,kanti,kanti2,kanti3,3D1,3D2,Panotopoulos:2017yoe,
Panotopoulos:2016wuu,Fernando:2004ay,coupling,chinos,Ahmed:2016lou} and references therein.
Now the QNM are determined requiring that the first coefficient of the second condition vanishes, i.e. $C_+ = 0$. The purely ingoing wave physically means that nothing can escape from the horizon, while the purely outgoing wave corresponds to the requirement that no radiation is incoming from infinity. We thus obtain an infinite set of discrete complex numbers $\omega=\omega_R + \omega_I i$ called the quasinormal frequencies of the black hole. Given the time dependence of the probe scalar field $\Phi \sim e^{-i \omega t}$, it is clear
that unstable modes correspond to $\omega_I > 0$, while stable modes correspond to $\omega_I < 0$. The real part of the mode $\omega_R$ determines the period of the oscillation, $T=2 \pi/\omega_R$, while the imaginary part $|\omega_I|$ describes the decay of the fluctuation at a time scale $t_D=1/|\omega_I|$.

\section{QN spectrum of scale dependent charged BH in EpM theory}

As usual in Physics, obtaining exact analytical solutions of realistic problems is extremely hard, and very few cases are known to exist. Computing the QN spectrum of black holes is no exception, and it does not come as a surprise the fact that only in some special cases analytical expressions have been obtained \cite{cardoso2,exact,potential,ferrari,fernando}. In this work we adopt the well-known from standard Quantum Mechanics WKB approximation \cite{wkb1,wkb2}, which is very popular and has been applied extensively to the literature. For an incomplete list see e.g. \cite{paper1,paper2,paper3} and for more recent works \cite{paper4,paper5,paper6,paper7,paper8} and references therein.

Just to fix the notation, the QN frequencies are given by
\begin{equation}
\omega^2 = V_0+(-2V_0'')^{1/2} \Lambda(n) - i \nu (-2V_0'')^{1/2} [1+\Omega(n)]
\end{equation}
where $n=0,1,2...$ is the overtone number, $\nu=n+1/2$, $V_0$ is the maximum of the effective potential, $V_0''$ is the second derivative of the effective potential evaluated at the maximum, while $\Lambda(n), \Omega(n)$ are complicated expressions of $\nu$ and higher derivatives of the potential evaluated at the maximum, and can be seen e.g. in \cite{paper2,paper7}. Here we have used the Wolfram Mathematica \cite{wolfram} code with WKB at any order from one to six presented in \cite{code}.

We work with the lapse metric function presented in section 2
\begin{equation}
f(r) = \frac{4 G_0 Q_0^2}{3 r (1+r \epsilon)^3}-\frac{G_0 M_0 (r^3 \epsilon^2+3 r^2 \epsilon + 3 r)}{3 r (1+r \epsilon)^3}
\end{equation}
where $G_0,Q_0,M_0$ are the classical values of the gravitational coupling, the electric charge and the mass of the BH respectively. 
From now on for simplicity we drop the index 0. In the following we fix $G=1=M$, and we give emphasis on the effect of the running on the spectrum, rather than computing as many frequencies as possible, considering 3 values of the angular momentum, namely $m=0,1,2$ and $n=0$, and 3 values of the running
parameter, namely $\epsilon=0, 0.05, 0.1$. Note that contrary to the standard Reissner-Nordstr{\"o}m BH as well as to the charged BH in four-dimensional EpM theory, where there are an inner and an outer event horizon (and also extremal BHs), here there is a single event horizon.

We summarize our results in the tables \ref{tab:1}, \ref{tab:2} and \ref{tab:3} below, while figures \ref{fig:3} and \ref{fig:4} show the real and the imaginary part of the frequencies respectively versus the electric charge $Q$ for the fundamental mode $m=0=n$. Each table corresponds to a certain $(m,n)$ pair,
in which we show the spectrum for several values of the electric charge and for 3 values of the running parameter, namely $\epsilon=0$ (classical case), $\epsilon=0.05$ and $\epsilon=0.1$. In figures \ref{fig:3} and \ref{fig:4}  the 3 curves correspond to these 3 values of the running parameter. We see that for $m=1$ and $m=2$ both the real and the imaginary part of the frequencies is positive, and thus the modes are unstable. The frequencies increase with the angular momentum and slightly with the running parameter, and decrease with the electric charge, but they always remain positive. In addition, for $m=0$ we observe the following features: a) In the classical case $\epsilon=0$ the modes are stable, while b) when we consider running of the couplings the modes are stable up to a certain value of the electric charge $Q_*$, after which the frequencies acquire a positive imaginary part. This special value is $Q_*=2.13$ for $\epsilon=0.05$ and $Q_*=1.39$ for $\epsilon=0.1$.
In previous works \cite{instability1,instability2} a similar behaviour was observed, albeit in a completely different context. In particular, in \cite{instability1} it was shown that modes are stable for $m \leq 0$ and unstable for $m > 0$, while in \cite{instability2} the imaginary part of the modes change sign for a certain value of the graviton mass.

\begin{figure}[ht!]
\centering
\includegraphics[width=\linewidth]{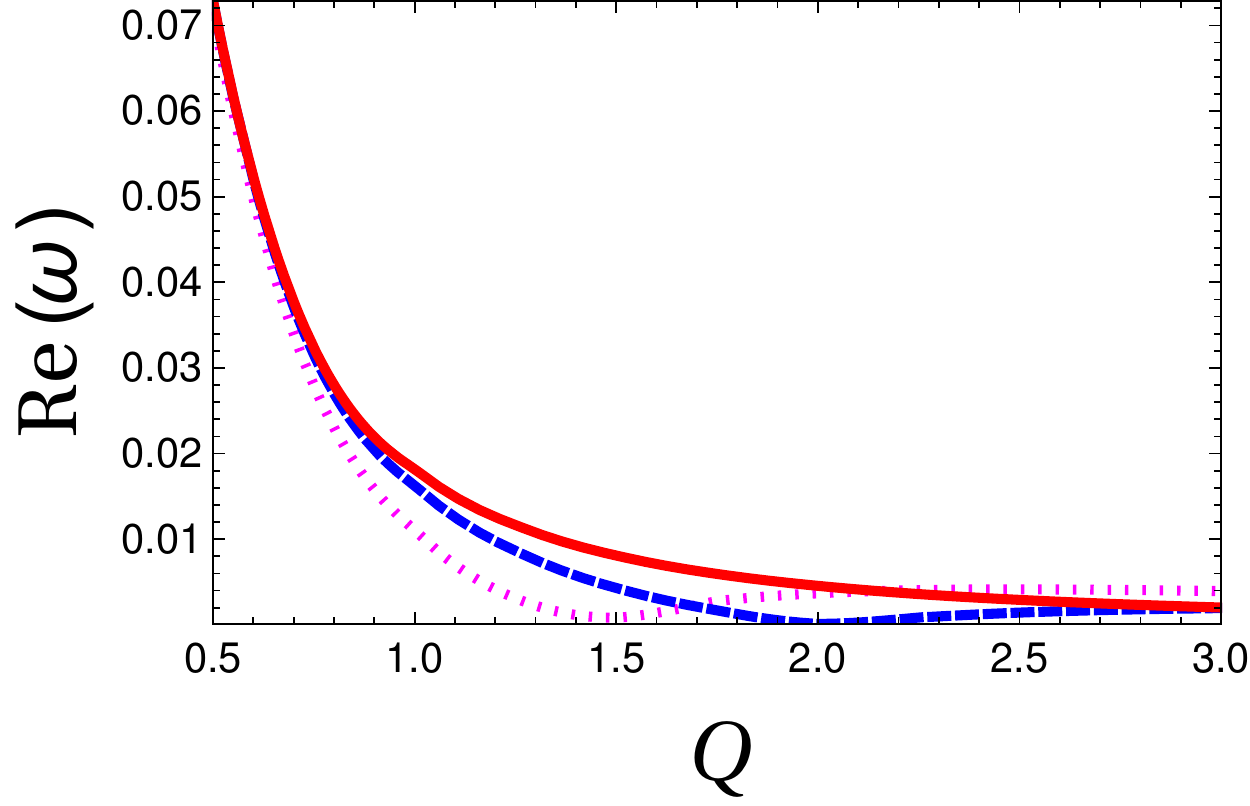}
\caption{\label{fig:3}
Re$(\omega)$ as a function of the electric charge $Q$ for the fundamental mode $m=0=n$ taking $G=M=1$ for: i) $\epsilon = 0$, (solid red line), ii) $\epsilon = 0.05$  (dashed blue line) and iii) $\epsilon = 0.1$  (dotted magenta line).
}
\end{figure}

\begin{figure}[ht!]
\centering
\includegraphics[width=\linewidth]{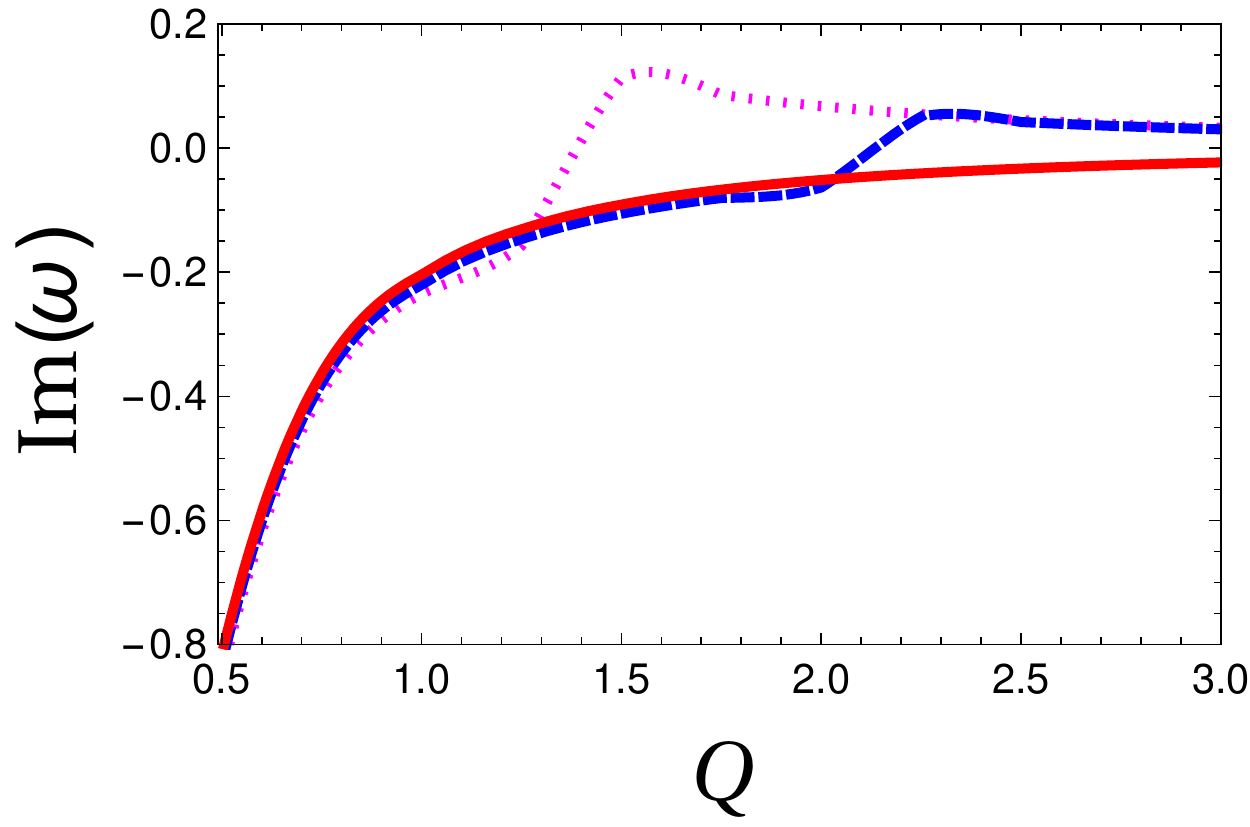}
\caption{\label{fig:4}
Im$(\omega)$ as a function of the electric charge $Q$ for the fundamental mode $m=0=n$ taking $G=M=1$ for: i) $\epsilon = 0$, (solid red line), ii) $\epsilon = 0.05$  (dashed blue line) and iii) $\epsilon = 0.1$  (dotted magenta line).
}
\end{figure}


\begin{table*}
\centering
  \caption{\label{tab:1} Quasinormal modes for the fundamental mode $m=0=n$ assuming different values of the electric charge $Q$ for $\epsilon=0, 0.05, 0.1$.}
  \begin{tabular}{c c c c}
  \hline
$Q$ & $ \epsilon=0 $ & $ \epsilon=0.05 $ & $\epsilon=0.1$ \\
\hline
0.50 & 0.0729\, -0.8168 i & 0.0728\, -0.8331 i  & 0.0716\, -0.8488 i \\
0.75 & 0.0324\, -0.3632 i & 0.0316\, -0.3791 i  & 0.0280\, -0.3954 i \\
1.00 & 0.0182\, -0.2042 i & 0.0163\, -0.2205 i  & 0.0111\, -0.2345 i \\
1.25 & 0.0117\, -0.1307 i & 0.0087\, -0.1465 i  & 0.0031\, -0.1573 i \\
1.50 & 0.0081\, -0.0908 i & 0.0043\, -0.1055 i  & 0.0008\, +0.1136 i \\
1.75 & 0.0059\, -0.0667 i & 0.0017\, -0.0801 i  & 0.0028\, +0.0863 i \\
2.00 & 0.0046\, -0.0511 i & 0.0001\, -0.0631 i  & 0.0037\, +0.0679 i \\
2.25 & 0.0036\, -0.0404 i & 0.0008\, +0.0511 i  & 0.0041\, +0.0551 i \\
2.50 & 0.0029\, -0.0327 i & 0.0014\, +0.0424 i  & 0.0041\, +0.0456 i \\
2.75 & 0.0024\, -0.0270 i & 0.0018\, +0.0357 i  & 0.0041\, +0.0385 i \\
3.00 & 0.0020\, -0.0227 i & 0.0020\, +0.0306 i  & 0.0039\, +0.0331 i \\
\hline
\end{tabular}
\end{table*}

%
%
%
%

\begin{table*}

\centering

\caption{\label{tab:2}Quasinormal modes assuming different values of the charge $Q$ for $m=1$ and for $\epsilon=0, 0.05, 0.1$.}
  \begin{tabular}{c c c c}
  \hline
$Q$ & $ \epsilon=0 $ & $ \epsilon=0.05 $ & $\epsilon=0.1$ \\
\hline
0.10  & 84.6627\, +126.4660 i  & 84.6552\, +126.5200 i    & 84.6457\, +126.5780 i \\
0.20  & 21.1639\, +31.6191 i   & 21.1654\, +31.6604 i     & 21.1668\, +31.7029 i \\
0.30  & 9.4067\, +14.0522 i    & 9.4060\, +14.0974 i      & 9.4087\, +14.1403 i \\
0.40  & 5.2913\,  +7.9043 i    & 5.2923\,  +7.9482 i      & 5.2954\,  +7.9933 i \\
0.50  & 3.3863\, +5.0590 i     & 3.3882\, +5.1027 i       & 3.3928\, +5.1492 i \\
0.60  & 2.3515\, +3.5132 i     & 2.3542\, +3.5574 i       & 2.3608\, +3.6050 i \\
0.70  & 1.7278\, +2.5809 i     & 1.7310\, +2.6261 i       & 1.7398\, +2.6745 i \\
0.80  & 1.3228\, +1.9762 i     & 1.3269\, +2.0218 i       & 1.3379\, +2.0711 i \\
0.90  & 1.0452\, +1.5614 i     & 1.0503\, +1.6076 i       & 1.0635\, +1.6578 i \\
1.00  & 0.8465\, +1.2648 i     & 0.8527\, +1.3115 i       & 0.8681\, +1.3623 i \\
2.00  & 0.2116\, +0.3162 i     & 0.2284\, +0.3665 i       & 0.2557\, +0.4168 i \\
5.00  & 0.0339\, +0.0506 i     & 0.0584\, +0.0959 i       & 0.0794\, +0.1306 i \\
10.00 & 0.0085\, +0.0127 i     & 0.0284\, +0.0468 i       & 0.0422\, +0.0695 i \\
\hline
\end{tabular}
\end{table*}

%
%
%

\begin{table*}

\centering

  \caption{\label{tab:3}Quasinormal modes assuming different values of the charge $Q$ for $m=2$ and for $\epsilon=0, 0.05, 0.1$.}
  \begin{tabular}{c c c c}
  \hline
$Q$ & $ \epsilon=0 $ & $ \epsilon=0.05 $ & $\epsilon=0.1$ \\
\hline

0.10  & 1286.10\, +2080.04 i   & 1286.11\, +2080.08 i  & 1286.11\, +2080.12 i \\
0.20  & 321.526\, +520.010 i   & 321.528\, +520.048 i  & 321.529\, +520.087 i \\
0.30  & 142.900\, +231.116 i   & 142.902\, +231.154 i  & 142.904\, +231.194 i \\
0.40  & 80.3815\, +130.0030 i  & 80.3832\, +130.0420 i & 80.3852\, +130.0830 i \\
0.50  & 51.4441\, +83.2016 i   & 51.4460\, +83.2412 i  & 51.4481\, +83.2845 i \\
0.60  & 35.7251\, +57.7789 i   & 35.7270\, +57.8193 i  & 35.7292\, +57.8651 i \\
0.70  & 26.2470\, +42.4498 i   & 26.2490\, +42.4912 i  & 26.2515\, +42.5399 i \\
0.80  & 20.0954\, +32.5006 i   & 20.0974\, +32.5432 i  & 20.1004\, +32.5952 i \\
0.90  & 15.8778\, +25.6795 i   & 15.8800\, +25.7233 i  & 15.8835\, +25.7790 i \\
1.00  & 12.8610\, +20.8004 i   & 12.8633\, +20.8456 i  & 12.8676\, +20.9054 i \\
2.00  & 3.2153\, +5.2001 i     & 3.2218\, +5.2665 i    & 3.2478\, +5.3792 i \\
5.00  & 0.5144\, +0.8320 i     & 0.5691\, +0.9705 i    & 0.6676\, +1.1467 i \\
10.00 & 0.1286\, +0.2080 i     & 0.2140\, +0.3675 i    & 0.2943\, +0.5043 i \\
\hline
\end{tabular}
\end{table*}

\section{Conclusions}

To summarize, in the present work we have studied the stability against scalar perturbations of a three-dimensional charged BH in EpM theory assuming running couplings. We have considered the case $k = 3/4$ for which the electromagnetic stress energy tensor is traceless. Starting from the wave equation for a massless scalar field we have obtained a Schr{\"o}dinger-like equation with an effective potential, and we have adopted the sixth order WKB approximation to obtain the quasinormal modes. Our numerical results have been summarized in tables, and we have shown graphically the dependence of the real and the imaginary part of the spectrum on the electric charge for 3 values of the running parameter. Our findings show that i) all modes corresponding to $m > 0$ are unstable, and ii) the fundamental mode without running is stable, while with running it is stable only up to a certain value of the electric charge, which is determined.


\begin{acknowledgments}
We would like to thank V. Cardoso for communications.
The work of A.R. was supported by the CONICYT-PCHA/Doctorado Nacional/2015-21151658.
G.P. thanks the Funda\c c\~ao para a Ci\^encia e Tecnologia (FCT), Portugal, for the
financial support to the Multidisciplinary Center for Astrophysics (CENTRA),  Instituto
Superior T\'ecnico,  Universidade de Lisboa, through the Grant No. UID/FIS/00099/2013.
\end{acknowledgments}



%
%


\begin{thebibliography}{99}
\bibitem{QG1} T.~Jacobson,
  Phys.\ Rev.\ Lett.\  {\bf 75} (1995) 1260
  [gr-qc/9504004].

\bibitem{QG2} A.~Connes,
  Commun.\ Math.\ Phys.\  {\bf 182} (1996) 155
  [hep-th/9603053].

\bibitem{QG3} M.~Reuter,
  Phys.\ Rev.\ D {\bf 57} (1998) 971
  [hep-th/9605030].

\bibitem{QG4} C.~Rovelli,
  Living Rev.\ Rel.\  {\bf 1} (1998) 1
  [gr-qc/9710008].

\bibitem{QG5} R.~Gambini and J.~Pullin,
  Phys.\ Rev.\ Lett.\  {\bf 94} (2005) 101302
  [gr-qc/0409057].

\bibitem{QG6} A.~Ashtekar,
  New J.\ Phys.\  {\bf 7} (2005) 198
  [gr-qc/0410054].

\bibitem{QG7} P.~Nicolini,
  Int.\ J.\ Mod.\ Phys.\ A {\bf 24} (2009) 1229
  [arXiv:0807.1939 [hep-th]].

\bibitem{QG8} P.~Horava,
  Phys.\ Rev.\ D {\bf 79} (2009) 084008
  [arXiv:0901.3775 [hep-th]].

\bibitem{QG9} E.~P.~Verlinde,
  JHEP {\bf 1104} (2011) 029
  [arXiv:1001.0785 [hep-th]].

\bibitem{hawking1} S.~W.~Hawking,
  Nature {\bf 248} (1974) 30.

\bibitem{hawking2} S.~W.~Hawking,/
  Commun.\ Math.\ Phys.\  {\bf 43} (1975) 199
   Erratum: [Commun.\ Math.\ Phys.\  {\bf 46} (1976) 206].

\bibitem{wheeler} T.~Regge and J.~A.~Wheeler,
  Phys.\ Rev.\  {\bf 108} (1957) 1063.

\bibitem{zerilli1} F.~J.~Zerilli,
  Phys.\ Rev.\ Lett.\  {\bf 24} (1970) 737.

\bibitem{zerilli2} F.~J.~Zerilli,
  Phys.\ Rev.\ D {\bf 2} (1970) 2141.

\bibitem{zerilli3} F.~J.~Zerilli,
  Phys.\ Rev.\ D {\bf 9} (1974) 860.

\bibitem{moncrief} V.~Moncrief,
  Phys.\ Rev.\ D {\bf 12} (1975) 1526.

\bibitem{teukolsky} S.~A.~Teukolsky,
  Phys.\ Rev.\ Lett.\  {\bf 29} (1972) 1114.

\bibitem{monograph} S.~Chandrasekhar,
  ``The mathematical theory of black holes,''
  OXFORD, UK: CLARENDON (1985) 646 P.

\bibitem{ligo1} B.~P.~Abbott {\it et al.} [LIGO Scientific and Virgo Collaborations],
  Phys.\ Rev.\ Lett.\  {\bf 116} (2016) no.6,  061102
[arXiv:1602.03837 [gr-qc]].

\bibitem{ligo2} B.~P.~Abbott {\it et al.} [LIGO Scientific and Virgo Collaborations],
  Phys.\ Rev.\ Lett.\  {\bf 116} (2016) no.24,  241103
[arXiv:1606.04855 [gr-qc]].

\bibitem{ligo3} B.~P.~Abbott {\it et al.} [LIGO Scientific and VIRGO Collaborations],
  Phys.\ Rev.\ Lett.\  {\bf 118} (2017) no.22,  221101
[arXiv:1706.01812 [gr-qc]].

\bibitem{review1} K.~D.~Kokkotas and B.~G.~Schmidt,
  Living Rev.\ Rel.\  {\bf 2} (1999) 2
[gr-qc/9909058].

\bibitem{review2} E.~Berti, V.~Cardoso and A.~O.~Starinets,
  Class.\ Quant.\ Grav.\  {\bf 26} (2009) 163001
  [arXiv:0905.2975 [gr-qc]].

\bibitem{CS} A.~Achucarro and P.~K. Townsend,
{\em Phys. Lett.}, vol.~B180, p.~89, 1986.

\bibitem{witten1} E.~Witten, ``{(2+1)-Dimensional Gravity as an Exactly Soluble System},'' {\em
  Nucl. Phys.}, vol.~B311, p.~46, 1988.

\bibitem{witten2} E.~Witten,
  arXiv:0706.3359 [hep-th].

\bibitem{btz1} M.~Banados, C.~Teitelboim and J.~Zanelli,
  Phys.\ Rev.\ Lett.\  {\bf 69} (1992) 1849
  [hep-th/9204099].

\bibitem{btz2} M.~Banados, M.~Henneaux, C.~Teitelboim and J.~Zanelli,
  Phys.\ Rev.\ D {\bf 48} (1993) 1506
   Erratum: [Phys.\ Rev.\ D {\bf 88} (2013) 069902]
  [gr-qc/9302012].

\bibitem{BH1} O.~Gurtug, S.~H.~Mazharimousavi and M.~Halilsoy,
  Phys.\ Rev.\ D {\bf 85} (2012) 104004
[arXiv:1010.2340 [gr-qc]].

\bibitem{BH2} M.~Hassaine and C.~Martinez,
  Class.\ Quant.\ Grav.\  {\bf 25} (2008) 195023
[arXiv:0803.2946 [hep-th]].

\bibitem{Rincon:2017ypd} \'A.~Rinc\'on, B.~Koch and I.~Reyes,
  J.\ Phys.\ Conf.\ Ser.\  {\bf 831}, no. 1, 012007 (2017)
  [arXiv:1701.04531 [hep-th]].

\bibitem{Rincon:2017goj} \'A.~Rinc\'on, E.~Contreras, P.~Bargueño, B.~Koch, G.~Panotopoulos and A.~Hernández-Arboleda,
  Eur.\ Phys.\ J.\ C {\bf 77} (2017) no.7,  494
  [arXiv:1704.04845 [hep-th]].

\bibitem{Koch:2014joa}
  B.~Koch, P.~Rioseco and C.~Contreras,
  Phys.\ Rev.\ D {\bf 91}, no. 2, 025009 (2015)
  [arXiv:1409.4443 [hep-th]].

\bibitem{angel} B.~Koch, I.~A.~Reyes and \'A.~Rinc\'on,
Class.\ Quant.\ Grav.\  {\bf 33} (2016) no.22,  225010
  [arXiv:1606.04123 [hep-th]].

\bibitem{Rincon:2017ayr} \'A.~Rinc\'on and B.~Koch,
  arXiv:1705.02729 [hep-th].

\bibitem{Bonanno:2000ep} A.~Bonanno and M.~Reuter,
  Phys.\ Rev.\ D {\bf 62}, 043008 (2000)
  [hep-th/0002196].

\bibitem{Bonanno:2006eu} A.~Bonanno and M.~Reuter,
  Phys.\ Rev.\ D {\bf 73}, 083005 (2006)
  [hep-th/0602159].

\bibitem{Reuter:2006rg}
  M.~Reuter and E.~Tuiran,
  hep-th/0612037.

\bibitem{Reuter:2010xb} M.~Reuter and E.~Tuiran,
  Phys.\ Rev.\ D {\bf 83}, 044041 (2011)
  [arXiv:1009.3528 [hep-th]].

\bibitem{Falls:2012nd} K.~Falls and D.~F.~Litim,
  Phys.\ Rev.\ D {\bf 89}, 084002 (2014)
  [arXiv:1212.1821 [gr-qc]].

\bibitem{Cai:2010zh} Y.~F.~Cai and D.~A.~Easson,
  JCAP {\bf 1009}, 002 (2010)
  [arXiv:1007.1317 [hep-th]].

\bibitem{Becker:2012js} D.~Becker and M.~Reuter,
  JHEP {\bf 1207}, 172 (2012)
  [arXiv:1205.3583 [hep-th]].

\bibitem{Becker:2012jx}
  D.~Becker and M.~Reuter,
  arXiv:1212.4274 [hep-th].

\bibitem{Koch:2013owa}
  B.~Koch and F.~Saueressig,
  Class.\ Quant.\ Grav.\  {\bf 31}, 015006 (2014)
  [arXiv:1306.1546 [hep-th]].

\bibitem{Koch:2013rwa}
  B.~Koch, C.~Contreras, P.~Rioseco and F.~Saueressig,
  Springer Proc.\ Phys.\  {\bf 170}, 263 (2016)
  [arXiv:1311.1121 [hep-th]].

\bibitem{Ward:2006vw}
  B.~F.~L.~Ward,
  Acta Phys.\ Polon.\ B {\bf 37}, 1967 (2006)
  [hep-ph/0605054].

\bibitem{Burschil:2009va}
  T.~Burschil and B.~Koch,
  Zh.\ Eksp.\ Teor.\ Fiz.\  {\bf 92}, 219 (2010)
  [JETP Lett.\  {\bf 92}, 193 (2010)]
  [arXiv:0912.4517 [hep-ph]].

\bibitem{Falls:2010he}
  K.~Falls, D.~F.~Litim and A.~Raghuraman,
  Int.\ J.\ Mod.\ Phys.\ A {\bf 27}, 1250019 (2012)
  [arXiv:1002.0260 [hep-th]].

\bibitem{Koch:2014cqa}
  B.~Koch and F.~Saueressig,
  Int.\ J.\ Mod.\ Phys.\ A {\bf 29}, no. 8, 1430011 (2014)
  [arXiv:1401.4452 [hep-th]].

\bibitem{Bonanno:2016dyv}
  A.~Bonanno, B.~Koch and A.~Platania,
  arXiv:1610.05299 [gr-qc].

\bibitem{Reuter:2003ca}
  M.~Reuter and H.~Weyer,
  Phys.\ Rev.\ D {\bf 69}, 104022 (2004)
  [hep-th/0311196].

\bibitem{Koch:2010nn}
  B.~Koch and I.~Ramirez,
  Class.\ Quant.\ Grav.\  {\bf 28}, 055008 (2011)
  [arXiv:1010.2799 [gr-qc]].

\bibitem{Domazet:2012tw}
  S.~Domazet and H.~Stefancic,
  Class.\ Quant.\ Grav.\  {\bf 29}, 235005 (2012)
  [arXiv:1204.1483 [gr-qc]].

\bibitem{Contreras:2016mdt}
  C.~Contreras, B.~Koch and P.~Rioseco,
  J.\ Phys.\ Conf.\ Ser.\  {\bf 720}, no. 1, 012020 (2016).

\bibitem{Percacci:2016arh}
  R.~Percacci and G.~P.~Vacca,
  Eur.\ Phys.\ J.\ C {\bf 77}, no. 1, 52 (2017)
  [arXiv:1611.07005 [hep-th]].

\bibitem{Contreras:2013hua}
  C.~Contreras, B.~Koch and P.~Rioseco,
  Class.\ Quant.\ Grav.\  {\bf 30}, 175009 (2013)
  [arXiv:1303.3892 [astro-ph.CO]].

\bibitem{Koch:2015nva}
  B.~Koch and P.~Rioseco,
  Class.\ Quant.\ Grav.\  {\bf 33}, 035002 (2016)
  [arXiv:1501.00904 [gr-qc]].

\bibitem{col1} C. Doran,  A. Lasenby,  S. Dolan,  and  I.  Hinder,  Phys.
Rev. D {\bf 71}, 124020 (2005).

\bibitem{col2} S. Dolan, C. Doran, and A. Lasenby, Phys. Rev. D {\bf 74}, 064005 (2006).

\bibitem{col3} L. C. B. Crispino, E. S. Oliveira, A. Higuchi, and G. E. A. Matsas,
Phys. Rev. D {\bf 75}, 104012 (2007).

\bibitem{col4} S.  R.  Dolan,  Classical  Quantum  Gravity {\bf 25}, 235002 (2008).

\bibitem{col5} L. C. B. Crispino, S. R. Dolan, and E. S. Oliveira,
Phys. Rev. Lett. {\bf 102}, 231103 (2009).

\bibitem{col6} L. C. B. Crispino, S. R. Dolan, and E. S. Oliveira,
Phys. Rev. D {\bf 79}, 064022 (2009).

\bibitem{col7} L. B. Crispino, A. Higuchi, and E. S. Oliveira,
Phys. Rev. D {\bf 80}, 104026 (2009).

\bibitem{kanti} P.~Kanti and J.~March-Russell,
  Phys.\ Rev.\ D {\bf 66} (2002) 024023
  [hep-ph/0203223].

\bibitem{kanti2} P.~Kanti, T.~Pappas and N.~Pappas,
  Phys.\ Rev.\ D {\bf 90} (2014) no.12,  124077
  [arXiv:1409.8664 [hep-th]].

\bibitem{kanti3} T.~Pappas, P.~Kanti and N.~Pappas,
  Phys.\ Rev.\ D {\bf 94} (2016) no.2,  024035
  [arXiv:1604.08617 [hep-th]].

\bibitem{3D1} D.~Birmingham, I.~Sachs and S.~Sen,
  Phys.\ Lett.\ B {\bf 413} (1997) 281
  [hep-th/9707188].

\bibitem{3D2} Y.~S.~Myung,
  Mod.\ Phys.\ Lett.\ A {\bf 18} (2003) 617
  [hep-th/0201176].

\bibitem{Panotopoulos:2017yoe} G.~Panotopoulos and \'A.~Rinc\'on,
  Phys.\ Rev.\ D {\bf 96}, no. 2, 025009 (2017)
[arXiv:1706.07455 [hep-th]].

\bibitem{Panotopoulos:2016wuu} 
  G.~Panotopoulos and \'A.~Rinc\'on,
  Phys.\ Lett.\ B {\bf 772}, 523 (2017)
  doi:10.1016/j.physletb.2017.07.014
  [arXiv:1611.06233 [hep-th]].

\bibitem{Fernando:2004ay} S.~Fernando,
  Gen.\ Rel.\ Grav.\  {\bf 37} (2005) 461,
  [hep-th/0407163].

\bibitem{coupling} L.~C.~B.~Crispino, A.~Higuchi, E.~S.~Oliveira and J.~V.~Rocha,
  Phys.\ Rev.\ D {\bf 87} (2013) 104034
  [arXiv:1304.0467 [gr-qc]].

\bibitem{chinos} Y.~Liu and J.~L.~Jing,
  Chin.\ Phys.\ Lett.\  {\bf 29} (2012) 010402.

\bibitem{Ahmed:2016lou} J.~Ahmed and K.~Saifullah,
  arXiv:1610.06104 [gr-qc].

\bibitem{cardoso2} V.~Cardoso and J.~P.~S.~Lemos,
  Phys.\ Rev.\ D {\bf 63} (2001) 124015
[gr-qc/0101052].

\bibitem{exact} D.~Birmingham,
  Phys.\ Rev.\ D {\bf 64} (2001) 064024
[hep-th/0101194].

\bibitem{potential} G.~Poschl and E.~Teller,
  Z.\ Phys.\  {\bf 83} (1933) 143.

\bibitem{ferrari} V.~Ferrari and B.~Mashhoon,
  Phys.\ Rev.\ D {\bf 30} (1984) 295.

\bibitem{fernando} S.~Fernando,
  Phys.\ Rev.\ D {\bf 77} (2008) 124005
  [arXiv:0802.3321 [hep-th]].

\bibitem{wkb1} S.~Iyer and C.~M.~Will,
  Phys.\ Rev.\ D {\bf 35} (1987) 3621.

\bibitem{wkb2} R.~A.~Konoplya,
  Phys.\ Rev.\ D {\bf 68} (2003) 024018
[gr-qc/0303052].

\bibitem{paper1} S.~Iyer,
  Phys.\ Rev.\ D {\bf 35} (1987) 3632.

\bibitem{paper2} K.~D.~Kokkotas and B.~F.~Schutz,
  Phys.\ Rev.\ D {\bf 37} (1988) 3378.

\bibitem{paper3} E.~Seidel and S.~Iyer,
  Phys.\ Rev.\ D {\bf 41} (1990) 374.

\bibitem{paper4} V.~Santos, R.~V.~Maluf and C.~A.~S.~Almeida,
  Phys.\ Rev.\ D {\bf 93} (2016) no.8,  084047
[arXiv:1509.04306 [gr-qc]].

\bibitem{paper5} S.~Fernando and C.~Holbrook,
  Int.\ J.\ Theor.\ Phys.\  {\bf 45} (2006) 1630
  doi:10.1007/s10773-005-9024-9
  [hep-th/0501138].

\bibitem{paper6} J.~L.~Blázquez-Salcedo, F.~S.~Khoo and J.~Kunz,
  arXiv:1706.03262 [gr-qc].

\bibitem{paper7} S.~K.~Chakrabarti,
  Gen.\ Rel.\ Grav.\  {\bf 39} (2007) 567
[hep-th/0603123].

\bibitem{paper8} R.~Konoplya,
  Phys.\ Rev.\ D {\bf 71} (2005) 024038
[hep-th/0410057].

\bibitem{wolfram}
\url{http://www.wolfram.com}

\bibitem{code} R.~A.~Konoplya and A.~Zhidenko,
  Phys.\ Rev.\ D {\bf 81} (2010) 124036
  [arXiv:1004.1284 [hep-th]].

\bibitem{instability1} S.~L.~Detweiler,
  Phys.\ Rev.\ D {\bf 22} (1980) 2323.

\bibitem{instability2} R.~Brito, V.~Cardoso and P.~Pani,
  Phys.\ Rev.\ D {\bf 88} (2013) no.2,  023514
  [arXiv:1304.6725 [gr-qc]].
\end{thebibliography}
\end{document}